\documentclass[%
 reprint,
 showkeys,
 amsmath,amssymb,
 aps,
prc,
]{revtex4-1}

\usepackage{graphicx}
\usepackage{dcolumn}
\usepackage{bm}
\usepackage{hyperref}

\begin{document}

\preprint{APS/123-QED}

\title{Fractality in momentum space: a signal of criticality in nuclear collisions}

\author{N.~G.~Antoniou}
 \email{nantonio@phys.uoa.gr}
\author{N.~Davis}%
 \email{ndavis@phys.uoa.gr}
\author{F.~K.~Diakonos}
 \email{fdiakono@phys.uoa.gr}
\affiliation{%
 Department of Physics, University of Athens, Athens, Greece
}%



\begin{abstract}
We show that critical systems of finite size develop a fractal structure in momentum space with anomalous dimension given in terms of the isotherm critical exponent $\delta$ of the corresponding infinite system. The associated power laws of transverse momentum correlations, in high-energy nuclear collisions, provide us with a signature of a critical point in strongly interacting matter according to the laws of QCD. 
\end{abstract}


\maketitle



Thermal systems of large but finite size develop a fractal structure in coordinate space, as they approach a critical point in the phase diagram \cite{Stinchcombe}. This geometrical property of critical systems is the manifestation of finite-size scaling and it is valid for length scales smaller than the correlation length $\xi$, $|\delta \bm{x}| \ll \xi$ \cite{Lesne}. The self-similar structure in question has a characteristic fractal dimension, $d_F = \frac{\delta d}{\delta+1}$, which depends on the topological dimension ($d$) and the isotherm critical exponent $\delta$ \cite{Antoniou1998}. An important issue related to this distinct property of a critical system is whether a similar geometrical structure is developed in momentum space \cite{Bialas}. In particular, for systems generated in high-energy collisions, momentum correlations may become crucial observables for the detection of critical fluctuations, once the above issue is resolved. To this end and in order to be specific, we consider in what follows a typical system belonging to the 3d Ising universality class in which the order parameter is a density function $\rho(\bm{x})$. In ordinary matter, $\rho(\bm{x})$ may refer to liquid-gas phase transitions \cite{Ma}, whereas in strongly interacting matter (quark matter) the order parameter is associated with the QCD critical point \cite{order} and may be identified with the density of scalars (chiral condensates) the formation of which as quasi-particles is likely to occur in high-energy nuclear collisions through the vacuum excitation ($\sigma$-field quanta).

The quasi-particle density $\rho(\bm{x})$ is given by the field operator $\phi(\bm{x})$ as: $\rho(\bm{x}) = \phi^*(\bm{x}) \phi(\bm{x})$ and its Fourier transform, with an obvious notation, reads:
\begin{equation}
 \rho_{\bm{k}} = \int d^3{\bm{x}}\, \mathrm{e}^{-i \bm{k} \bm{x}}\, \rho(\bm{x}) \qquad \text{or} \qquad \rho_{\bm{k}} = \sum_{\bm{p}}\, a^{*}_{\bm{p}}\, a^{}_{\bm{p}+\bm{k}}
 \label{rhoFourier}
\end{equation}
where the creation and annihilation operators $(a^{*}_{\bm{p}}, a^{}_{\bm{p}})$ are introduced with the standard integral: $\phi(\bm{x}) = \int d^3{\bm{p}}\, \mathrm{e}^{-i \bm{p} \bm{x}}\, a_{\bm{p}}$. The quantity $\langle \rho_{\bm{k} = \bm{0}} \rangle$ gives the average multiplicity of quasi-particles in a multiparticle production process:
\begin{equation}
 \langle \rho_{\bm{k} = \bm{0}} \rangle = \int d^3{\bm{x}} \langle \rho(\bm{x}) \rangle = \sum_{\bm{p}} \langle a^{*}_{\bm{p}}\, a^{}_{\bm{p}} \rangle = \sum_{\bm{p}} \langle n_{\bm{p}} \rangle
 \label{rhoKappa0}
\end{equation}
where $n_{\bm{p}}$ is the eigenvalue of the quasi-particle number operator for momentum $\bm{p}$. The correlation integral $\langle \rho_{\bm{k}}^*\, \rho_{\bm{k}} \rangle$ is written, with the help of the representation (\ref{rhoFourier}) as follows:

\begin{align}
\begin{split}
 \langle \rho_{\bm{k}}^*\, \rho_{\bm{k}} \rangle &= V\, \int d^3{\bm{x}}\, \mathrm{e}^{-i \bm{k} \bm{x}}\, \langle \rho(\bm{x})\, \rho(\bm{0}) \rangle \qquad \text{or}\\
 \langle \rho_{\bm{k}}^*\, \rho_{\bm{k}} \rangle &= \sum_{\bm{p},\bm{p'}} \langle a^{*}_{\bm{p}+\bm{k}}\, a^{}_{\bm{p}}\, a^{*}_{\bm{p'}}\, a^{}_{\bm{p'}+\bm{k}} \rangle
 \end{split}
 \label{rhoKappa}
\end{align}

The Fourier integral in eq.(\ref{rhoKappa}) is valid for a system of large but finite volume V, under the assumption that the dominant contribution comes from the domain far from the surface of the boundary (bulk) where spatial translation symmetry holds. On the other hand, the terms with $\bm{p} \neq \bm{p'}$ in the summation (\ref{rhoKappa}) vanish in any momentum state $\left| \Psi \right\rangle$ of the many-body system under consideration:
\begin{align}
\begin{split}
\left\langle \Psi \right|&\, a^{*}_{\bm{p}+\bm{k}}\, a^{}_{\bm{p}}\, a^{*}_{\bm{p'}}\, a^{}_{\bm{p'}+\bm{k}}  \left| \Psi \right\rangle = 0 \quad \text{for } \bm{p} \neq \bm{p'}\\
\left| \Psi \right\rangle& = \left| \ldots n_{\bm{p}}\, n_{\bm{p}+\bm{k}}\, n_{\bm{p'}}\, n_{\bm{p'}+\bm{k}} \ldots  \right\rangle
\end{split}
\end{align}%
as a result of a successive action of the operators $(a,a^*)$. Therefore, the summation in eq.(\ref{rhoKappa}) is simplified as follows:
\begin{align}
 \begin{split}
  \langle \rho_{\bm{k}}^*\, \rho_{\bm{k}} \rangle &= \sum_{\bm{p}} \langle a^{*}_{\bm{p}+\bm{k}}\, a^{}_{\bm{p}}\, a^{*}_{\bm{p}}\, a^{}_{\bm{p}+\bm{k}} \rangle \quad \text{or}\\
  \langle \rho_{\bm{k}}^*\, \rho_{\bm{k}} \rangle &= \sum_{\bm{p}} \langle n_{\bm{p}+\bm{k}} \, (n_{\bm{p}} + 1) \rangle
 \end{split}
 \label{rhoKappaP}
\end{align}

In a critical system of finite size, the power-law of the correlator in configuration space, $\langle \rho(\bm{x})\,  \rho(\bm{0})\rangle \sim |\bm{x}|^{d_F-d}$, at scales close to the correlation length $\xi$, as dictated by the fractal structure with $d_F = \frac{\delta d}{\delta+1}$ \cite{Antoniou1998}, implies a power-law singularity of the Fourier transform, $\langle \rho_{\bm{k}}^*\, \rho_{\bm{k}} \rangle \sim  |\bm{k}|^{-d_F}$, in the limit of small $\bm{k}$ \cite{Falconer}. This singularity is transfered to the right-hand side of eq.(\ref{rhoKappaP}):
\begin{equation}
 \lim_{|\bm{k}| \to \bm{0}} \langle \rho_{\bm{k}}^*\, \rho_{\bm{k}} \rangle = \lim_{|\bm{k}| \to \bm{0}} \sum_{\bm{p}} \langle n_{\bm{p}+\bm{k}} \, n_{\bm{p}} \rangle + \langle n \rangle
\end{equation}
where the summation represents the pair correlation in momentum space, $\rho_{12} (\bm{k}) = \sum_{\bm{p}} \langle n_{\bm{p}+\bm{k}} \, n_{\bm{p}} \rangle$, and $\langle n \rangle = \sum_{\bm{p}} \langle n_{\bm{p}} \rangle$. Finally:
\begin{align}
 \begin{split}
  \lim_{|\bm{k}| \to \bm{0}}  \langle \rho_{\bm{k}}^*\, \rho_{\bm{k}} \rangle &= \lim_{|\bm{k}| \to \bm{0}} \rho_{12} (\bm{k}) \qquad \text{or}\\
  \lim_{|\bm{k}| \to \bm{0}}  \rho_{12} (\bm{k}) &\sim |\bm{k}|^{-d_F}
 \end{split}
 \label{powerLawK}
\end{align}
namely, the singular part of the Fourier transform $\langle \rho_{\bm{k}}^*\, \rho_{\bm{k}} \rangle$ in eq.(\ref{rhoKappa}) coincides with the singular part of the correlator in momentum space, $\rho_{12} (\bm{k})$. In fact, the power-law (\ref{powerLawK}) reveals a fractal structure in momentum space with dimension $\tilde{d}_F = d - d_F$ or $\tilde{d}_F = \frac{d}{\delta+1}$. This structure provides us with a tool for the detection of a critical point belonging to the universality class $(d,\delta)$ in experiments of multiparticle production. In particular, owing to this effect, the phenomenology of the QCD critical point in high-energy nuclear collisions \cite{Stephanov} can be founded on a solid basis employing the method of intermittency which has been designed in order to study fractals in momentum space \cite{BialasP}. The above arguments, when applied to a critical system produced in relativistic nuclear collisions, maintain their validity only if we project the system onto the transverse momentum plane of the collision. In fact, in this plane the Fourier transform in eq.(\ref{rhoKappa}) remains at work whereas along the longitudinal direction the validity of a power-law requires extra assumptions associated with the space-time nature of the rapidity variable \cite{DeWolf}. The fractal dimension in transverse momentum space is $\tilde{d}_F = \frac{2}{\delta+1}$, under the assumption of approximate cylindrical geometry in which longitudinal and transverse motion are disentangled. As a result, we claim that the signature of the QCD critical point in these experiments is a strong 2d intermittency effect in the transverse momentum plane of the production process either of sigma quasi-particles, in the form of $\pi^+\pi^-$ pairs \cite{Antoniou2005}, or of net protons \cite{Antoniou2006}. Quantitatively, the signal of the critical point is given by the intermittency index (exponent) which is predicted by the laws of QCD (universality class). 

In particular, owing to the fact that the QCD critical point belongs to the 3d Ising universality class \cite{order} in which $\delta \simeq 5$, the fractal dimension in transverse momentum space is, according to the discussion above, $\tilde{d}_F \simeq \frac{1}{3}$. As a result, the intermittency efffect of protons, linked to critical fluctuations, leads to a power-law, $F_2(M) \sim M^\frac{5}{3}$ ($M \gg 1$), of the second factorial moment $F_2$ as a function of $M^2$, the number of 2d cells in momentum space \cite{Antoniou2006}. In a real critical system produced in relativistic ion collisions the power-law in configuration space will hold between two scales, i.e. for $R_{min} < \vert \vec{x} \vert < R_{max}$. As a consequence, according to the previous discussion, the counter part in momentum space is a power-law with the exponent obtained above, holding in the range $k_{min} < \vert \vec{k} \vert < k_{max}$ with $k_{min} \approx 1/R_{max}$ and $k_{max} \approx 1/R_{min}$. As discussed in the introduction $R_{max}$ is determined by the correlation length in configuration space, which, at the critical point, is of the order of the size $V_d^{1/d}$ of the system (where $V_d$ is the associated volume). Furthermore, $R_{min}$ is estimated in \cite{Antoniou1998} to $R_{min} \sim \left(\frac{V_d}{\langle n \rangle}\right)^{1/d}$. In order to observe the fractal structure formed in transverse momentum space for a fireball freezing out at the QCD critical point, the upper scale $k_{max} \approx \left(\frac{\langle n \rangle}{V_d}\right)^{1/d}$ should be much greater than the experimental momentum resolution $\delta k$. This condition is ideally fulfilled for medium size nuclei at high (but not too high to enter into the crossover regime) beam energy.

In summary, we have shown that despite the fact that the correlation integral $\langle \rho_{\bm{k}}^*\, \rho_{\bm{k}} \rangle$ in a multiparticle process does not coincide with the pair correlation in momentum space, $\rho_{12}(\bm{k}) \neq \langle \rho_{\bm{k}}^*\, \rho_{\bm{k}} \rangle$, they both share the same singular behavior for small $\bm{k}$, in a critical system. This property leads to a fractal structure in momentum space with a well prescribed fractal dimension in terms of the universality class of the critical point.

\begin{acknowledgements}
 We wish to thank the members of the NA49 experiment and in particular Stanislaw Mrowczynski for fruitful discussions. One of us (\mbox{N. Davis}) wishes to thank the Research Institute ``Demokritos'', Greece, for financial support.
\end{acknowledgements}

\end{document}